# Vogls: a Fast Interactive Full-timing Simulator for Pre-silicon Power Side-Channel Analysis

Gijs Burghoorn
*Radboud University*
Nijmegen, The Netherlands
gijs.burghoorn@ru.nl

Ileana Buhan
*Radboud University*
Nijmegen, The Netherlands
ileana.buhan@ru.nl

Lejla Batina
*Radboud University*
Nijmegen, The Netherlands
lejla.batina@ru.nl

***Abstract*—Designing hardware circuits resistant to side-channel attacks increasingly relies on simulation to predict device leakage before fabrication. Current functional verification simulators are designed for extended correctness-checking runs and are ill-suited for producing large numbers of short trace collections with slight input variants needed for side-channel analysis. We present Vogls: an open-source Verilog simulator built for side-channel analysis, that is the first simulator to combine compiled-code performance, full-timing simulation, and fine-grained control over the simulation state. Vogls simulates a timing annotated gate-level AES design 5.9 times faster than Icarus Verilog and is only 30% slower than Verilator on a PicoRV32 RTL design while also offering a more accurate timing model. Vogls provides a Python interface that allows simulations to be paused, forked, inspected, and mutated around points-of-interest without modifying the hardware design and thereby preserving timing and leakage characteristics. Furthermore, a trace-collection workflow runs setup code once and forks at the point-of-interest, eliminating the need to re-simulate from reset for every trace. A case study demonstrates Vogls on a differential power analysis attack that successfully recovers the key at RTL, GTL and full-timing GTL abstraction levels.**

## I. Introduction

Software and hardware may leak information through power consumption [1], electromagnetic emanations [2] and timing characteristics [3]. *Side-Channel Analysis* (SCA) determines whether there is leakage and finds the source to mitigate it. Side-channel attacks rely on physical measurements of power consumption or electromagnetic emanations and have proven to be a realistic threat [4]. Consequently, government bodies impose requirements for side-channel resistance [5]. *Test-Vector Leakage Assessment* (TVLA) [6] is widely used in SCA to detect side-channel leakage, while *Differential Power Analysis* (DPA) [1] exploits such leakage to recover secrets by statistically comparing sets of power consumption traces.

Designing countermeasures for side-channel attacks is an iterative process. For software targets, iteration is cheap and new versions can be generated and immediately tested. In contrast, for hardware targets, each iteration requires fabrication, incurring significant cost, turnaround time, and engineering effort. Consequently, pre-silicon hardware SCA tools that evaluate countermeasures before fabrication have become an active research area. These tools use formal methods or simulation to detect leakage (determining whether leakage exists) or to locate leakage (identifying the responsible components). Formal methods [7] provide soundness guarantees within their threat model, but scale poorly to larger designs and suffer from false positives. Simulation provides a workflow similar to physical measurements and is already widely used for hardware design. However, it provides no soundness guarantees.

Academic pre-silicon power SCA tools [8], [9] that utilize simulation follow a three-step pattern. First, they use or modify a *Hardware Description Language* (HDL) simulator to capture signal traces. Depending on the abstraction level and timing effects, a tool may use cycle-accurate or full-timing simulation. The signal traces are written to *Value Change Dump* (VCD) [10], *Switch-Activity Interchange Format* (SAIF) [11] or *Fast Signal Trace* (FST) [12] files. Second, the tools convert signal traces into synthetic power traces using a power model such as Hamming Weight, Hamming Distance or Non-Linear power estimation [13]. Third, they perform standard power SCA on the synthetic power traces. Although simulated traces are typically noiseless, which causes attacks to need fewer traces, protected or large designs may still need thousands to millions of simulated traces. Liu et al. note that for a large number of traces, these simulations take a prohibitive amount of time and signal trace files grow excessively large, making the storage and conversion steps costly [14].

This paper presents the *Versatile Open Gate Level Simulator* (Vogls): a toolchain for full-timing digital hardware simulation built for SCA. Vogls will be open-sourced upon acceptance of this paper and provides the following contributions over the state-of-the-art:

- Vogls compiles timed simulations to C, allowing faster full-timing simulation than other open-source simulators. Consequently, Vogls reduces trace-collection time, making previously infeasible SCA workflows practical.
- Vogls allows forking, inspecting, and altering simulation states. A workflow can therefore run setup logic once, then fork, mutate, and simulate only around the point-of-interest. Because the design itself is never modified, its timing and leakage characteristics are preserved.
- Vogls implements the Verilog language, which is widely used in hardware design. Consequently, Vogls fits into existing workflows without design modifications.

Additionally, this paper introduces the *Vogls Intermediate Representation* (VIR), which provides a unified target covering simulation semantics from cycle-accurate through full-timing, and we present a case study that illustrates using Vogls to perform a successful pre-silicon DPA attack.

The paper is structured as follows. Section II provides the necessary background and discusses related work. Then, Section III covers the design and implementation of Vogls. Afterward, Section IV quantitatively compares Vogls against other simulators and presents a DPA attack using Vogls. Section V covers the current limitations and future directions of Vogls. Finally, Section VI concludes the paper.

## II. Background and Related Work

This section describes background and prior work that influenced Vogls' design. The section first discusses simulation techniques and other hardware simulators. Then, it covers circuit representations. Lastly, it discusses several pre-silicon SCA tools.

### A. Simulation Techniques

Hardware designs are written in Verilog, VHDL or other *Hardware Description Languages* (HDLs), and HDL simulators vary along five main axes: abstraction level, timing model, logic model, execution strategy and interfacing strategy.

a) *Abstraction Level:* Hardware designs go through stages in the digital design flow, including *Register-Transfer Level* (RTL), technology-mapping, layout, placement and routing. Each stage produces a circuit representation with a specific abstraction level. A *Gate-Level* (GTL) netlist refers to such circuit representation after technology-mapping. As the design flow advances, simulations of these representations are more precise but more resource intensive. Liu et al. outline that different classes of side-channel vulnerabilities are exposed at different abstraction levels [14]. Consequently, a simulator fitted to SCA ideally supports multiple abstraction levels.

b) *Timing model:* As hardware descriptions advance through synthesis, estimating and simulating finer-grained timing behavior of circuits, such as propagation delay or clock skew, becomes feasible and important. To accurately model timing, a simulator needs to track sub-cycle circuit states. Alternatively, a simulator may choose to be cycle-accurate, meaning it is accurate only at clock cycle boundaries, which generally allows much faster and less resource-consuming simulation. Mangard et al. illustrate that timing effects, e.g. glitches, may circumvent SCA countermeasures and thus pre-silicon SCA needs to consider full-timing models [15].

c) *Logic model:* Digital circuits in simulation are modeled using different logic models. The two most common models are two-value logic, which includes logical low `0` and logical high `1` , and four-value logic, which additionally includes the high impedance `Z` and unknown `X` . Designs may depend on four-value logic to model uninitialized state and buses. Similarly, four-value logic may increase the accuracy of timing simulations. However, simulations of simpler models are less resource intensive and standard SCA leakage models, such as Hamming Weight or Hamming Distance, do not have a natural definition for four-value logic. Consequently, SCA trace generation needs to support both two-value and four-value logic.

d) *Execution strategy:* A simulator can interpret the design or compile a design to native code. Compiling a design trades some time up front for later simulation performance. Consequently, interpreted execution is better for iteration and short or one-off runs, and compiled execution is better for long or repeated runs. SCA commonly requires thousands or millions of traces on the same design with slight input variations. Therefore, SCA trace collection prefers compiled execution with reusable state.

e) *Interfacing strategy:* Simulators also differ in the level of control they offer users. Some, such as Icarus Verilog [16], mostly allow interaction at the start. Others, such as Verilator [17], expose the simulation as a library that can be programmatically controlled. Most simulators implement the *Verilog Procedural Interface* (VPI) to allow for some observation and mutation. VPI is used by verification frameworks such as CocoTB [18], but does not allow copying or forking the simulation state. Forking is useful for SCA trace-collection since simulations typically contain setup logic that can take thousands of cycles before reaching a point-of-interest that may only take a few cycles. With forking, this setup runs once and the resulting state is reused and modified between traces. Furthermore, there are *adaptive* (or *online*) side-channel attacks [19]. These attacks choose the next query based on the traces or outputs from previous queries. A simulator for these attacks must therefore allow inspecting traces and adapting subsequent queries interactively.

We extract three criteria for an SCA focused HDL simulator. A simulator should be **(1) efficient**: capture enough signal traces for SCA in reasonable time, **(2) interactive**: pause, fork, inspect and mutate the simulation state during runs, and **(3) versatile**: run across different abstraction levels, timing models and logic models.

### B. Current HDL Simulators

Given the above criteria, we survey commonly used simulators, also noting whether source-code is publicly available. Table I summarizes the comparison.

Verilator [17] is an open-source, cycle-accurate, two-value logic simulator that transpiles designs to C++. This C++ code can be used as a library, making Verilator **(2) interactive**. It is widely used and regarded as **(1) efficient** and scalable. However, it does not support full-timing simulations or four-value logic and thus is not **(3) versatile**.

Icarus Verilog [16] is an open-source Verilog simulator that interprets a stack-based intermediate representation. It is **(3) versatile** as it supports four-value logic, full-timing simulations and several abstraction levels. It is used as a reference simulator for Verilog designs, but satisfies neither **(1) efficient** nor **(2) interactive** as Icarus Verilog is relatively slow compared to other simulators and makes it hard to control the simulation state.

Widely used commercial simulators are generally **(1) efficient**, **(3) versatile**, but none are **(2) interactive**. For these simulators the source-code is not available, except CVC [20], which provides source-code upon request. For the remainder of the paper, we restrict our comparison to non-commercial open-source simulators, both for reproducibility and because commercial license terms generally prevent publishing benchmarks.

TABLE I
OVERVIEW OF HDL SIMULATORS

| | Open | Efficient | Interactive | Versatile |
|---|---|---|---|---|
| Vogls | ✓ | ✓ | ✓ | ✓ |
| Verilator | ✓ | ✓ | ✓ | ✗ |
| Icarus Verilog | ✓ | ✗ | ✗ | ✓ |
| CVC | ✗[1] | ✓ | ✗ | ✓ |
| Commercial Simulators | ✗ | ✓ | ✗ | ✓ |

To our knowledge, no existing Verilog simulator combines (1) **efficient**, (2) **interactive**, and (3) **versatile**. Open-source tools like Yosys [21], OpenROAD [22], Verilator [17] and Surfer [23] have formed an open-source toolbox in academia covering synthesis, place-and-route, cycle-accurate simulation and waveform viewing. Vogls fills a gap in this toolbox: fast, scriptable, full-timing simulation.

Since Vogls is a HDL simulator, it could be used for functional verification, but it specifically focuses on facilitating the pre-silicon SCA workflow. Vogls does this by focusing on the three previously mentioned criteria, implementing the Hamming Weight and Hamming Distance power models, and providing a scriptable Python interface.

### C. Circuit Representations

Many intermediate representations exist for digital circuits supporting synthesis or functional analysis. *Multi-Level Intermediate Representation* (MLIR) [24] provides primitives and forms a family of dialects for (high-level) synthesis. It brings ideas from the LLVM compiler infrastructure [25] to hardware synthesis. *Flexible Intermediate Representation for RTL* (FIRRTL) [26] targets RTL and focuses on synthesis transformation. *Low-Level Hardware Description* (LLHD) [27] models hardware at several abstraction levels to provide synthesis transforms. It includes some temporal information, but does not target full-timing as it does not support `specify` blocks or *Standard Delay Format* (SDF) [28] files. The CIRCT project builds on MLIR to provide modular, reusable components. Both FIRRTL and LLHD now have corresponding dialects in the CIRCT project.

MLIR, FIRRTL and LLHD focus on synthesizable circuits. Gate-level and timing annotated simulation usually rely on non-synthesizable semantics in the Verilog language such as delay control, specify blocks and timing checks. The semantics of these timing-accurate simulations are Verilog-specific and impact the signal-propagation behavior especially with respect to signal glitches.

### D. Simulation-based Pre-silicon SCA Tools

AMASIVE [29] was among the first tools proposed for pre-silicon hardware SCA. It does not use simulation but instead converts an RTL model into a graph representation and proposes a set of graph algorithms that can detect vulnerabilities. RTL-PSC [30] uses RTL simulations to generate synthetic power traces. The authors use *Kullback-Leibler* (KL) divergence to rank block-level leakage across the RTL simulations, GTL simulations and *Field-Programmable Gate Array* (FPGA) measurements. They demonstrate high correlation in rankings at different abstraction levels. ACA [8] utilizes GTL simulations to rank cells according to their leakage. It uses a commercial power estimation tool to additionally consider electrical effects in their ranking. PATCH [9] provides a tool to both detect and mitigate side-channel leakage. It also uses commercial power estimation tools to determine especially leaky parts of the circuit. Telescope [14] compares pre-silicon side-channel leakage at several abstraction levels. They illustrate that leakage appearing at one abstraction level may not appear at other abstraction levels. Buhan et al. provide a wider survey of the tools used for software and hardware SCA [31].

Pre-silicon SCA tools are judged according to their *efficacy*: how well a tool correctly identifies side-channel leakage. Specifically, an effective tool minimizes false-positives and false-negatives for specific targets. Since the full space of possible side-channel attacks is unknown, it is difficult to quantify these rates. In practice, these rates are therefore estimated relative to a specific attack scenario. Furthermore, efficacy can only be evaluated in relation to a specific abstraction level, timing model and leakage model. Since Vogls is model- and attack-agnostic, its efficacy depends on the specific analysis.

## III. DESIGN OF VOGLS

Vogls is an open-source tool implemented in Rust whose development is driven by the needs of SCA research. This section gives a high-level overview of Vogls, then we describe the intermediate representation and its significance, and finally we describe how simulation state is managed.

### A. Overview of Vogls

Fig. 1 contains a block diagram that illustrates how Vogls turns a Verilog design into a bytecode or compiled code. Vogls is structured as a typical programming language compiler. Verilog is tokenized, parsed, type-checked and then transformed into an intermediate representation. There are two notable differences from programming language compilers. First, Verilog needs elaboration, which occurs between parsing and type-checking and expands the module hierarchy. Second, Vogls uses a temporally-aware intermediate representation as discussed in Section III.B.

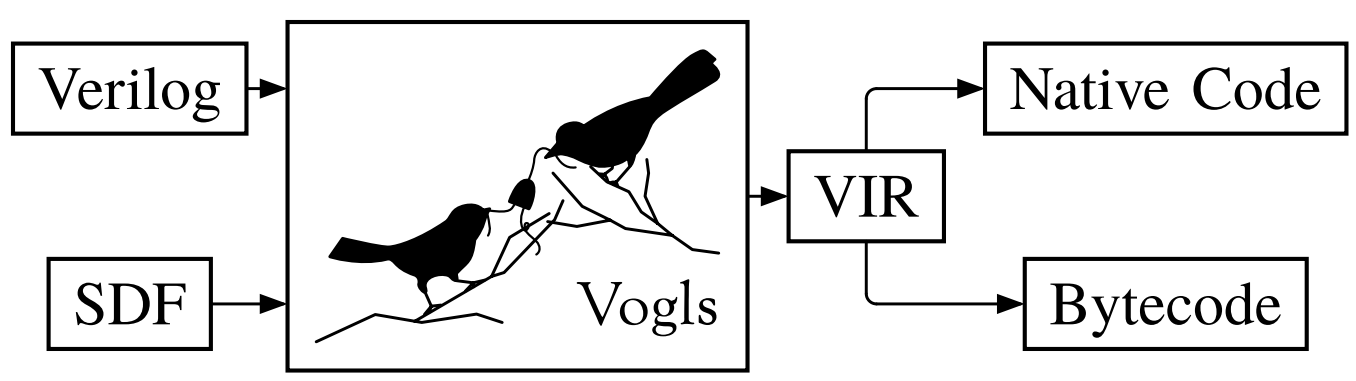

Fig. 1. A block diagram of Vogls taking Verilog to bytecode or native code.

Vogls optimizes the intermediate representation to make compilation and simulation faster and less resource-intensive while preserving observable behavior. Optimizations include standard compiler optimizations such as constant propagation, dead-code elimination, peephole optimization and common sub-expression elimination. However, there are also HDL-specific optimizations such as combining equivalent wires to reduce scheduler events and memory consumption.

[1]CVC's source-code is available upon request

The optimized intermediate representation gets converted to one of two execution targets. Vogls' bytecode interpreter allows for short runs and fast iteration. Alternatively, Vogls can output C source-code, which an external C compiler compiles into native processor instructions. This is typically better for repeated runs or large designs.

### *B. Vogls Intermediate Representation*

Vogls uses the *Vogls Intermediate Representation* (VIR) to create a unified target for a circuit's temporal and logic semantics and to simplify lowering, optimization and evaluation. VIR is inspired by *Low-Level Hardware Description* (LLHD) [27], but supports simulation-specific and non-synthesizable Verilog concepts. It balances a compact instruction set with specialized instructions to enable efficient evaluation. This section discusses VIR's semantics and design decisions.

The representation consists of a collection of processes that point to a basic block as their execution entry-point. Each basic block consists of instructions and a terminator. Instructions — for instance `%z = xor %l, %r` or `%z = prb $signal` — describe *Static Single-Assignment* (SSA) operations on variables (prefixed by an `%`) and signals (prefixed by a `$`). Each variable and signal has a fixed bit-width. Terminators — for instance `halt`, `branch %condition, <truthy>, <falsy>`, and `wait #5, <next>` — point to other basic blocks forming a *Control-Flow Graph* (CFG). VIR uses SSA and CFGs, which are both well-established in the compiler literature, to simplify analysis and transformations.

Semantically, processes execute in parallel but a process' instructions execute sequentially. A process may be *active*, *scheduled*, *listening* (after calling `watch`), or *done* (after calling `halt`). When a process is scheduled, it is in the *active* region, another region in the current timestamp, or scheduled for a future timestamp. When the active region is empty, all processes in the first non-empty region are moved to the active region. If all regions in the current timestamp are empty, all the processes from the next lowest timestamp are moved into the active region. When a signal is updated (with the `drv $signal, %value` instruction), all processes that are listening to `$signal` are placed in the active region.

Circuit timing annotations — such as `specify` blocks, wire delays or SDF annotations, which are typically provided by technology libraries — also get mapped to VIR. This is done by inserting processes that use the *Last Update Time* (`lupdt`) instruction and the `varwait` terminator. The `%t0 = lupdt $signal` instruction sets the `%t0` variable to the timestamp when `$signal` was last changed. Ending a basic block with `varwait %delay, <next>` sleeps the process for `%delay` time steps and jumps to the basic block labelled `next`. These concepts model delays as required for `specify` blocks and SDF annotations.

Vogls converts Verilog to VIR following Verilog scheduling semantics and a specification is available for VIR.

### *C. Separating State from Design*

Vogls separates a simulation's *design* from *state*. The design remains constant across runs and includes signal names, types, configuration, and compiled code. The state might change over the course of the simulation and includes the simulation time, signal values, listeners, update times and event regions.

The design is constant and shared between runs. Consequently, forking a simulation only requires copying the state, which is considerably less work than re-initializing the design. For SCA, such separation allows a trace-collection workflow to run setup logic once, then fork before the point-of-interest and simulate a short window with varied inputs.

For forks to execute truly independently, the entire state must be copied and the design must remain read-only. If either requirement is violated, forking would introduce subtle, unpredictable differences between runs. The same two properties also allow simulations to run concurrently without synchronization. After forking, users commonly alter signal values before resuming. Vogls wakes affected listeners to ensure mutations get propagated when the simulation continues.

## IV. Experiments

This section shows experiments performed with Vogls. The first section shows a performance comparison of Vogls against Verilator and Icarus Verilog. The second section discusses a case study of using Vogls for pre-silicon SCA.

### *A. Performance Comparison*

We compare Vogls' performance against two other popular open-source simulators: Verilator and Icarus Verilog. All experiments were run on an Intel Core i9-10900K CPU running at 3.70GHz with 64GB of RAM. We benchmark on functional-verification workloads rather than SCA-specific workflows, as these are easier to quantify and allow fairer comparison across simulators. Specifically, we focus on the time to prepare a design, which determines the iteration speed after making changes, and the time to simulate a design, which determines how many traces can be collected. Generally, longer preparation time allows for better simulation performance. Since Vogls supports both a bytecode interpreter and compiled native code execution strategies, we report results for both execution strategies.

Table II shows the results of three experiments aimed at providing a fair comparison of the performance of Vogls and the two main other open-source simulators. The experiments focus on designs that closely resemble SCA targets.

1) Executing RISC-V instructions on the RTL design of a PicoRV32 [32] to calculate the $1000^{\text{th}}$ prime number.
2) Asserting all outcomes for an RTL design of a masked Canright SBOX [33] for three given masks. The OpenTitan root-of-trust[2] uses a similar component for its AES encryption.
3) Transferring over UART and encrypting using AES [34] on a timing-annotated GTL design synthesized by Yosys using the ICE40 FPGA cell library.

[2] https://github.com/lowrisc/opentitan

TABLE II

Compilation and simulation time in seconds to (1) calculate the 1000th prime number on the PicoRV32, (2) verify all outcomes of a masked Canright SBOX for 3 different masks, and (3) perform a full-timing gate-level simulation of UART + AES encryption using an ICE40 FPGA netlist.

| | *(1) PicoRV32* | | *(2) Canright SBOX* | | *(3) UART + AES* | |
|---|---|---|---|---|---|---|
| **Simulator** | **Compile (s)** | **Simulate (s)** | **Compile (s)** | **Simulate (s)** | **Compile (s)** | **Simulate (s)** |
| *Verilator* | 7.332 ± 0.332 | 1.441 ± 0.011 | 7.042 ± 0.221 | 0.004 ± 0.001 | - | - |
| *Icarus Verilog* | 0.024 ± 0.010 | 104.081 ± 0.305 | 0.030 ± 0.001 | 0.290 ± 0.001 | 0.417 ± 0.002 | 76.164 ± 1.049 |
| *Vogls (Interpreter)* | 0.005 ± 0.000 | 48.869 ± 0.061 | 0.006 ± 0.000 | 2.423 ± 0.005 | 0.170 ± 0.004 | 58.456 ± 0.624 |
| *Vogls (Compiler)* | 1.929 ± 0.014 | 1.799 ± 0.001 | 3.381 ± 0.024 | 0.158 ± 0.000 | 93.167 ± 0.013 | 12.798 ± 0.042 |

To provide a fair comparison, all simulators run the entire simulation on a single thread each time. This comparison is inherently imperfect. Verilator does not implement Verilog simulation semantics, which is likely advantageous for its runtime. Icarus Verilog performs full 4-value logic simulation, which is disadvantageous when not needed.

As can be seen in Table II, for the PicoRV32 design compiled Vogls simulates in 1.799s versus 1.441s for Verilator. However, both compiled simulators are much faster than the interpreted simulators. For the Canright SBOX design, compiled Vogls simulates in 0.158s versus 0.004s for Verilator and 0.290s for Icarus Verilog. This difference is likely due to the design generating many small Verilog processes for which neither Vogls nor Icarus Verilog currently have optimizations. For the full-timing simulation containing a UART and an AES implementation, compiled Vogls needs roughly 1.5 minutes to compile the design. Afterwards Vogls simulates in 12.798 s versus 76.164s for Icarus Verilog. Verilator cannot run this experiment, as it lacks full-timing support.

### B. Case Study

As a demonstration of using Vogls for SCA, we present a case study of a DPA attack on the third design from Section IV.A. The attack is a well-documented last-round AES attack using the Hamming Distance power model. We will utilize Vogls' simulation forking to make our trace collection more efficient. Similar analysis can be performed with Icarus Verilog, but would require restarting for each simulation, a complex testbench that is difficult to iterate on, or interaction through VPI, which is complex and limited. This section walks through the case study step-by-step and abbreviates code for clarity, but a Python notebook with the source-code is also available. Afterwards, we present the attack results.

Our analysis starts by defining the top-level design that we will be using:

```
1 import vogls as vg
2 dgn = vg.Design('design.v')
3 state = dgn.initial_state()
```

In our design, we locate the signals of interest. In later steps, we will inspect and change these signals.

```
1 nrst = dgn.signals.resolve('nrst')
2 # ...
3 Dout = dgn.signals.resolve('Dout')
```

In this case study, our setup steps consist of resetting our design and waiting for the encryption core to be ready.

```
1 dgn.signals.set(state, nrst, "1'b0")
2 dgn.run_for(state, 2 * CYCLE)
3 dgn.signals.set(state, nrst, "1'b1")
4 dgn.run_for(state, 256 * CYCLE)
```

Afterwards, we fork our state, creating a snapshot that we can return to and reuse as the starting point for each trace.

```
1 new_state = state.fork()
```

With that new state, we perform an encryption by setting the data input and key input, running for a set amount of cycles and inspecting the data output.

```
1 k = # ... fixed target key
2 dgn.signals.set(new_state, Din, pt)
3 dgn.signals.set(new_state, Key, k)
4 dgn.run_for(new_state, 14 * CYCLE)
5 ct = dgn.signals.get(new_state, Dout)
```

To trace the signals and get the Hamming Distance at every simulation timestamp, we request a trace before running our encryption and extract the signal activity after.

```
1 t = dgn.trace(state)
2 # ... perform encryption
3 hd = t.extract().hamming_distance()
```

Combining all the steps above, the complete code to capture a trace is:

```
1 new_state = state.fork()
2 t = dgn.trace(new_state)
3 pt = # ... random 128-bit plaintext
4 dgn.signals.set(new_state, Din, pt)
5 dgn.signals.set(new_state, Key, k)
6 dgn.run_for(new_state, 14 * CYCLE)
7 ct = dgn.signals.get(new_state, Dout)
8 hd = t.extract().hamming_distance()
```

This trace collection code can be executed in a loop sequentially or distributed across a concurrent thread pool. Having described the workflow, we present the results of the DPA attack. The attack captures 30 000 traces utilizing the loop-body specified above for the RTL, GTL and full-timing

GTL models. The gate-level designs use the ICE40 cell library and require minor adjustments to the signal assignments.

Table III shows the minimum amount of traces needed to disclose all the correct key bytes, i.e. the *Minimum Traces to Disclosure* (MTD). The table also includes the time needed to capture 30 000 traces using 8 worker threads. As can be seen, both the MTD and time to capture traces increase from RTL to GTL to full-timing GTL.

TABLE III
MINIMUM TRACES TO DISCLOSURE (MTD) AND THE TIME NEEDED TO CAPTURE 30,000 TRACES FOR THE RTL, GTL AND FULL-TIMING GTL MODELS.

| **Abstraction** | **MTD** | **Capture Time** | **Simulations / Second** |
|---|---|---|---|
| *RTL* | 6 897 | 1.0s | 28672.3 |
| *GTL* | 13 154 | 48.2s | 622.1 |
| *Full-Timing GTL* | 18 341 | 101.1s | 296.7 |

Fig. 2 shows the rank of the slowest converging key-byte as a function of the number of traces. Consistent with Table III, more detailed models need more traces for full disclosure.

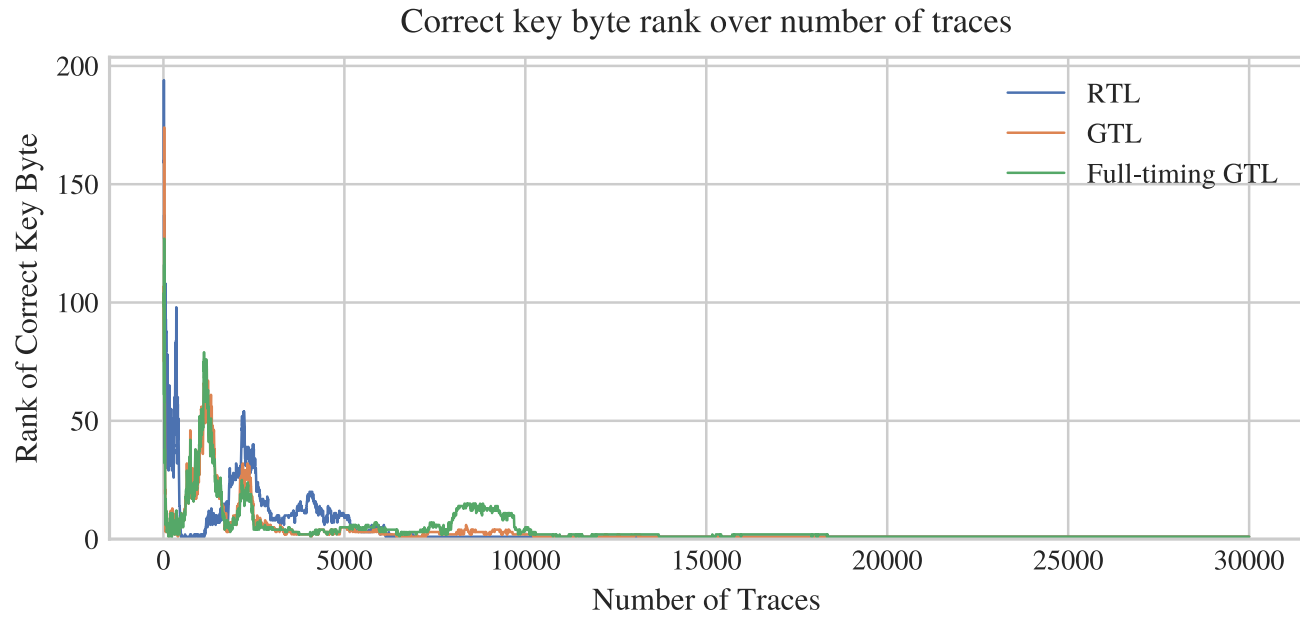

Fig. 2. Key-byte rank of the slowest-converging byte versus number of traces for RTL, GTL, and full-timing GTL models.

This case study illustrates how to use Vogls in a pre-silicon SCA workflow. It shows that Vogls enables practical SCA across the RTL, GTL and full-timing GTL abstraction levels.

## V. Discussion and Future Work

Although Vogls already illustrates what a SCA-oriented simulator looks like, it is at a considerably earlier stage of development than HDL simulators like Verilator and Icarus Verilog. We wish to emphasize that these simulators are mature, widely adopted, and the result of substantial engineering effort. Consequently, the aim of this paper is not to argue that we should move away from these established and great tools, but to highlight the advantages that a purpose-built design can offer and to motivate further exploration in this direction. This section lays out current limitations and future directions.

Development of Vogls is driven by SCA research, so several Verilog features — including driving strength, `trireg` , `real` and procedural continuous assignments — that are not required by those targets are not implemented. Similarly, concepts that are mostly employed for hardware verification, such as VPI, are not yet implemented. Consequently, Vogls is more lightweight and easier to extend while still supporting the necessary concepts for full-timing simulation and trace collection. The source-code repository includes a full overview of which Verilog concepts are supported. That said, we believe VIR's flexible execution model requires only incremental changes to support the entire Verilog standard since VIR already captures the required scheduling behavior.

Since HDL simulators form critical tools whose correctness is hard to establish, they should be verified with care. Vogls has a test-suite verifying implemented features in both interpreted and compiled mode. However, these tests mostly stem from bugs found during development. For Vogls to mature, it needs a broader testing strategy including end-to-end tests on realistic designs, systematic tests that explore many design configurations, and tracking the tested code-paths in Vogls.

In this paper, we show Vogls running a masked AES implementation, the PicoRV32 and an ICE40 gate-level design. During our research, we have also simulated the Ibex core[3] after using a tool to convert it from SystemVerilog to Verilog. The resulting Verilog comprises $18,000$ lines. Future research should aim to determine what Vogls' scalability limits are. One known limit is that the size of the generated C code and consequently the C compiler runtime grow prohibitive for large designs. This is also visible in experiment (3) of Table II.

Another future research direction is VIR optimizations to improve runtime and reduce the number of processes. Specifically, there are many structural and temporal optimizations that have been left unexplored until now. Optimization passes such as merging `always` blocks, pruning signals only used in a single process and fusing processes likely have a large impact on the compilation and simulation time.

We also believe that VIR has several potential uses outside Vogls. For example, formal verification tools that inspect timing related properties could utilize VIR as input representation. Furthermore, pre-silicon SCA tools that parse and process Verilog independently, such as [35], can utilize VIR.

A final future research direction is incorporating more SCA workflow steps into Vogls including pre-silicon TVLA and SCA metrics such as *Mutual Information* (MI) [36].

## VI. Conclusion

This paper describes Vogls: a new Verilog and gate-level simulator that focuses on criteria important to pre-silicon side-channel analysis. First, we outlined the Vogls Intermediate Representation (VIR) and how it allows for analysis and optimization of the design. Second, we described how Vogls manages the state to allow for forking of simulations and parallel simulations. Third, we showed that Vogls' runtime performance on an RTL design is only 30% slower than Verilator, which is the industry standard open-source cycle-accurate simulator, while being 5.9 times faster than Icarus Verilog on timing-annotated simulation. Fourth, we provided a case study of a DPA attack that successfully recovers an AES key using Vogls at three abstraction levels. By open-sourcing Vogls, we invite collaboration to provide better tools for pre-silicon side-channel analysis.

[3]https://github.com/lowRISC/ibex

# References

[1] P. Kocher, J. Jaffe, and B. Jun, “Differential Power Analysis,” in *Advances in Cryptology — CRYPTO’ 99*, M. Wiener, Ed., Berlin, Heidelberg: Springer, 1999, pp. 388–397. doi: 10.1007/3-540-48405-1_25.

[2] K. Gandolfi, C. Mourtel, and F. Olivier, “Electromagnetic Analysis: Concrete Results,” in *Cryptographic Hardware and Embedded Systems — CHES 2001*, Ç. K. Koç, D. Naccache, and C. Paar, Eds., Berlin, Heidelberg: Springer, 2001, pp. 251–261. doi: 10.1007/3-540-44709-1_21.

[3] P. C. Kocher, “Timing Attacks on Implementations of Diffie-Hellman, RSA, DSS, and Other Systems,” in *Advances in Cryptology — CRYPTO ’96*, N. Koblitz, Ed., Berlin, Heidelberg: Springer, 1996, pp. 104–113. doi: 10.1007/3-540-68697-5_9.

[4] T. Roche, “EUCLEAK.” [Online]. Available: https://eprint.iacr.org/2024/1380

[5] Federal Office for Information Security (BSI), “Side-Channel Resistance.” Accessed: Apr. 20, 2026. [Online]. Available: https://www.bsi.bund.de/EN/Themen/Unternehmen-und-Organisationen/Informationen-und-Empfehlungen/Kryptografie/Seitenkanalresistenz/seitenkanalresistenz.html?nn=916726

[6] B. J. Gilbert Goodwill, J. Jaffe, P. Rohatgi, and others, “A testing methodology for side-channel resistance validation,” in *NIST non-invasive attack testing workshop*, 2011, pp. 115–136.

[7] B. Gigerl, V. Hadzic, R. Primas, S. Mangard, and R. Bloem, “Coco: Co-Design and Co-Verification of Masked Software Implementations on CPUs,” presented at the 30th USENIX Security Symposium (USENIX Security 21), 2021, pp. 1469–1468. Accessed: Dec. 03, 2025. [Online]. Available: https://www.usenix.org/conference/usenixsecurity21/presentation/gigerl

[8] Y. Yao, T. Kathuria, B. Ege, and P. Schaumont, “Architecture Correlation Analysis (ACA): Identifying the Source of Side-channel Leakage at Gate-level.” Accessed: Dec. 02, 2025. [Online]. Available: https://eprint.iacr.org/2020/1192

[9] V. Samadi Bokharaie and A. Jahanian, “Power side-channel leakage assessment and locating the exact sources of leakage at the early stages of ASIC design process,” *The Journal of Supercomputing*, vol. 78, no. 2, pp. 2219–2244, Feb. 2022, doi: 10.1007/s11227-021-03927-w.

[10] I. S. Association and others, “IEEE standard for Verilog hardware description language (IEEE 1364-2005),” *http://standards. ieee. org/*, 2006.

[11] “IEEE Standard for Design and Verification of Low-Power Integrated Circuits.” Accessed: Apr. 14, 2026. [Online]. Available: http://ieeexplore.ieee.org/document/6521327/

[12] T. Hutt, “FST Format Specification.” [Online]. Available: https://blog.timhutt.co.uk/fst_spec/

[13] Y. Nasser, J. Lorandel, J.-C. Prévotet, and M. Hélard, “RTL to transistor level power modeling and estimation techniques for FPGA and ASIC: A survey,” *IEEE Transactions on Computer-Aided Design of Integrated Circuits and Systems*, vol. 40, no. 3, pp. 479–493, 2020.

[14] Z. Liu, A. Malnicof, A. Roy, and P. Schaumont, “Telescope: Top-Down Hierarchical Pre-silicon Side-channel Leakage Assessment in System-on-Chip Design,” in *Proceedings of the 20th ACM Asia Conference on Computer and Communications Security*, in ASIA CCS '25. New York, NY, USA: Association for Computing Machinery, Aug. 2025, pp. 1280–1293. doi: 10.1145/3708821.3736216.

[15] S. Mangard, N. Pramstaller, and E. Oswald, “Successfully attacking masked AES hardware implementations,” in *Proceedings of the 7th international conference on Cryptographic hardware and embedded systems*, in CHES'05. Berlin, Heidelberg: Springer-Verlag, Aug. 2005, pp. 157–171. doi: 10.1007/11545262_12.

[16] S. Williams and M. Baxter, “Icarus verilog: open-source verilog more than a year later,” *Linux Journal*, vol. 2002, no. 99, 2002, Accessed: Mar. 12, 2026. [Online]. Available: https://dl.acm.org/doi/10.5555/513581.513584

[17] W. Snyder, P. Wasson, D. Galbi, G. Lore, and et al, “Verilator.” Accessed: Mar. 11, 2026. [Online]. Available: https://github.com/verilator/verilator

[18] B. Rosser, “Cocotb: a Python-based digital logic verification framework.”

[19] L. Batina, Ł. Chmielewski, L. Papachristodoulou, P. Schwabe, and M. Tunstall, “Online template attacks,” *Journal of Cryptographic Engineering*, vol. 9, no. 1, pp. 21–36, Apr. 2019, doi: 10.1007/s13389-017-0171-8.

[20] S. Meyer, “CVC Verilog Compiler – Fast Complex Language Compilers Can be Simple.” Accessed: Mar. 12, 2026. [Online]. Available: http://arxiv.org/abs/1603.08059

[21] C. Wolf and J. Glaser, “Yosys - A Free Verilog Synthesis Suite.”

[22] T. Ajayi and D. Blaauw, “Openroad: Toward a self-driving, open-source digital layout implementation tool chain,” in *Proceedings of Government Microcircuit Applications and Critical Technology Conference*, 2019.

[23] F. Skarman, L. Klemmer, D. Große, O. Gustafsson, and K. Laeufer, “Surfer — An Extensible Waveform Viewer,” in *Computer Aided Verification*, R. Piskac and Z. Rakamarić, Eds., Cham: Springer Nature Switzerland, 2025, pp. 392–404. doi: 10.1007/978-3-031-98685-7_19.

[24] C. Lattner *et al.*, “MLIR: Scaling Compiler Infrastructure for Domain Specific Computation,” in *2021 IEEE/ACM International Symposium on Code Generation and Optimization (CGO)*, Feb. 2021, pp. 2–14. doi: 10.1109/CGO51591.2021.9370308.

[25] C. Lattner and V. Adve, “LLVM: a compilation framework for lifelong program analysis & transformation,” in *International Symposium on Code Generation and Optimization, 2004. CGO 2004.*, Mar. 2004, pp. 75–86. doi: 10.1109/CGO.2004.1281665.

[26] A. Izraelevitz *et al.*, “Reusability is FIRRTL ground: Hardware construction languages, compiler frameworks, and transformations,” in *2017 IEEE/ACM International Conference on Computer-Aided Design (ICCAD)*, Nov. 2017, pp. 209–216. doi: 10.1109/ICCAD.2017.8203780.

[27] F. Schuiki, A. Kurth, T. Grosser, and L. Benini, “LLHD: a multi-level intermediate representation for hardware description languages,” in *Proceedings of the 41st ACM SIGPLAN Conference on Programming Language Design and Implementation*, in PLDI 2020. New York, NY, USA: Association for Computing Machinery, Jun. 2020, pp. 258–271. doi: 10.1145/3385412.3386024.

[28] “IEEE Standard for Standard Delay Format (SDF) for the Electronic Design Process.” Accessed: Apr. 14, 2026. [Online]. Available: https://ieeexplore.ieee.org/document/972829/

[29] S. A. Huss, M. Stöttinger, and M. Zohner, “AMASIVE: An Adaptable and Modular Autonomous Side-Channel Vulnerability Evaluation Framework,” in *Number Theory and Cryptography: Papers in Honor of Johannes Buchmann on the Occasion of His 60th Birthday*, M. Fischlin and S. Katzenbeisser, Eds., Berlin, Heidelberg: Springer, 2013, pp. 151–165. doi: 10.1007/978-3-642-42001-6_12.

[30] M. He, J. Park, A. Nahiyan, A. Vassilev, Y. Jin, and M. Tehranipoor, “RTL-PSC: Automated Power Side-Channel Leakage Assessment at Register-Transfer Level,” in *2019 IEEE 37th VLSI Test Symposium (VTS)*, Apr. 2019, pp. 1–6. doi: 10.1109/VTS.2019.8758600.

[31] I. Buhan, L. Batina, Y. Yarom, and P. Schaumont, “SoK: Design Tools for Side-Channel-Aware Implementations,” in *Proceedings of the 2022 ACM on Asia Conference on Computer and Communications Security*, in ASIA CCS '22. New York, NY, USA: Association for Computing Machinery, May 2022, pp. 756–770. doi: 10.1145/3488932.3517415.

[32] “YosysHQ/picorv32.” Accessed: Apr. 21, 2026. [Online]. Available: https://github.com/YosysHQ/picorv32

[33] D. Canright and L. Batina, “A Very Compact "Perfectly Masked" S-Box for AES (corrected).” Accessed: Dec. 04, 2025. [Online]. Available: https://eprint.iacr.org/2009/011

[34] Tohoku University ECSIS Laboratory, “Cryptographic IP cores.” [Online]. Available: https://github.com/ECSIS-lab/Cryptographic-Hardware-IPs

[35] A. V. Lakshmy, C. Rebeiro, and S. Bhunia, “FORTIFY: Analytical Pre-Silicon Side-Channel Characterization of Digital Designs,” in *2022 27th Asia and South Pacific Design Automation Conference (ASP-DAC)*, Jan. 2022, pp. 660–665. doi: 10.1109/ASP-DAC52403.2022.9712551.

[36] L. Batina, B. Gierlichs, E. Prouff, M. Rivain, F.-X. Standaert, and N. Veyrat-Charvillon, “Mutual Information Analysis: a Comprehensive Study,” *Journal of Cryptology*, vol. 24, no. 2, pp. 269–291, Apr. 2011, doi: 10.1007/s00145-010-9084-8.